\documentclass[12pt]{article}
\newcommand{\be}{\begin{equation}}
\newcommand{\ee}{\end{equation}}
\newcommand{\bea}{\begin{eqnarray}}
\newcommand{\eea}{\end{eqnarray}}
\newcommand{\nn}{\nonumber}
\usepackage{amsmath}
\usepackage{amsfonts}
\usepackage{amssymb}
\begin{document}
\thispagestyle{empty}

\begin{center}
{\bf  \Large SPACE SYMMETRIES AND QUANTUM BEHAVIOUR
OF FINITE ENERGY CONFIGURATIONS IN $SU(2)$-GAUGE THEORY }\\
\vspace {10mm}
\setcounter{footnote}{0}
\renewcommand{\thefootnote}{\fnsymbol{footnote}}
A. V. Shurgaia$^{1,2}$\footnote{e-mail: avsh@rmi.acnet.ge}  and
H. J. W. M\"uller-Kirsten$^{1}$\footnote{e-mail: mueller1@physik.uni-kl.de}\\
\vspace{10mm}
{\footnotesize 1. Department of Physics,  University of Kaiserslautern, D-67653,
Kaiserslautern, Germany \\
2. Department of Theoretical Physics, A. Razmadze Mathematical
Institute \\ of the Georgian Academy of Sciences. \\ 1 M. Alexidze
str. 0193 Tbilisi, Georgia}
\end{center}

\begin{abstract}
The quantum properties of  localized finite energy solutions to classical Euler-Lagrange
equations are investigated using the method of collective coordinates. The perturbation theory
in terms of inverse powers of  the coupling constant $g$ is constructed, taking into account
the conservation laws of  momentum and angular momentum (invariance of the action with respect
to the group of motion $M(3)$ of 3-dimensional Euclidean space) rigorously in every order of perturbation theory.
\end{abstract}

coordinates, symmetry, perturbation theory. arXiv: hep-th/0605281
\newpage
{\bf 1. Introduction}
 \setcounter{footnote}{0}
\vspace{\baselineskip}

The discovery of extended objects in field theory has required the
development of methods for describing their quantum properties. One
of the most powerful methods is based on introducing collective
coordinates, which are related to the symmetry group of the physical
system. This method has a long history beginning from 1940
\cite{pauli} when it was applied to the model of fixed source. In
\cite{bog} this method has been formulated in the mathematically
rigorous and consistent form. In connection with  nonlinear
classical field theories this method has attracted attention and in
the papers of many authors it was developed for interacting systems
with arbitrary symmetries and applied to various quantum field
theoretical models with either the operator \cite{khrust} or path
integral formulation of the field theory \cite{gerv}. The general
covariant formulation of the method for the multicomponent  fields
of any kind on the basis of Dirac's generalized  Hamiltonian theory
\cite{dir} in the frame of the path integral approach to the quantum
theory is given in \cite{shur-93}. An alternative approach to the
covariant description of  field theoretical models has been given in
\cite{svesh}. The study of localized solutions in realistic models
has shown, that their quantum properties enable us to describe spin
structure \cite{shur-76,jackiw}. In order to achieve this the
collective coordinates must be introduced, which are related to the
full symmetry of the 3-dimensional space, that is the symmetry with
respect to the transformations of the group (named $M(3)$) of motion
of 3-dimensional  Euclidean space \cite{shur-76,vilen}.

In this paper we  describe the spin structure of pure $SU(2)$-gauge
theory, which admits classical localized finite energy regular
configurations \cite{'thooft} (a review on various possible such
configurations is given in \cite{actor}). This is the simplest model
that can be  studied. The more complicated gauge theories require to
describe the internal structure  also. There is a number of papers,
in which quantum properties of such solitonic configurations have
been investigated (see \cite{woo,actor} and references therein). The
important point in the study of gauge theories is, of course, the
gauge freedom. One tries always to introduce the gauge conditions in
order to cancel the unphysical degrees of freedom and then
constructs the quantum theory. In  ref. \cite{christ} a complete
analysis of possible gauges has been  performed showing either
positive or negative features of these. We only point out that the
difficulties regarding the possible gauges are related either to the
singularities of gauge transformations or to the nonlocal nature of
the reduced phase space and the violation of the spherical symmetry.
In the following   we proceed from full phase space in constructing
the Hamiltonian picture and pass to the quantum theory as usual. The
constraints will then be understood as operator equalities acting on
a corresponding Hilbert space. So the problem reduces to the
construction of the appropriate space of physical states, that is
the space of vectors which are simultaneously
 the eigenvectors of the Hamilton operator and conserved quantities and
 fulfills the requirements of the constraints.

\vspace{\baselineskip}

{\bf  2. Hamiltonian formalism}
\vspace{\baselineskip}

We consider nonabelian $SU(2)$-gauge theory in 4-dimensional
Minkowskian space with the action functional (the metric
($1,-1,-1,-1)$ will be used)\footnote{We use Greek letters for
vectors in Minkowskian space, the letters i,j,k... for vectors in
Euclidean space and a,b,c... for vectors in isotopic space.} \bea
S=\int d^4x{\cal L}(A_{\mu a},\partial_{\mu}A_{\nu a})=
-\frac{1}{4}\int d^4x F_{\mu\nu a}F^{\mu\nu}{\,}_{a}. \eea The field
strength tensor is:
 \bea F_{\mu\nu a}=\partial_{\mu}A_{\nu a}(x)-
\partial_{\nu}A_{\mu a}(x)+ g\varepsilon_{abc}A_{\mu b}(x)A_{\nu
c}(x), \eea with $A_{\mu a}(x)$ as a vector field belonging to the
adjoint representation of the gauge group $SU(2)$ and being a
solution of the set of differential equations: \bea
\partial^{\mu}F_{\mu\nu a}+g\varepsilon_{abc}A^{\mu}_{b}(x)F_{\mu\nu c}=0.
\eea We suggest that there is a space-localized finite energy static
solution. Proceeding from translational invariance of the theory we
suppose that the solution depends on parameters $x_{0i}$, which
represent the position of the localized solution at the initial
time.  Translations in space-time change the position of the
solution in such a way, that it could be  useful to parametrize the
solutions as $A_{\mu a}(x)=A_{\mu a}({\bf x},c_kt+x_{0k})$ which
means  that $\partial_0$ has to be replaced by $c_k\partial_k$. We
see that if $A_{\mu a}(x)=g^{-1}\tilde {A}_{\mu a}(x)$, then $\tilde
{A}_{\mu a}(x)$ does not depend on $g$. Let us pass to the
Hamiltonian formulation of the theory as we are going to work in the
Schr\"odinger picture. The Lagrangian of the model we are
considering is singular, that is the canonical variables fulfill
some constraint equations. We shall make use of the generalized
Hamiltonian formalism developed by Dirac \cite{dir} for such
physical systems for constructing the Hamilton's equations of
motion. Let us define  $A_{\mu a}(x)$ and their conjugate momenta
\bea \pi^{\mu}_{a}(x)= \frac{\partial{\cal L}}{\partial\dot{A}_{\mu
a}(x)}=F^{\mu 0}{\,}_{a} \eea as the set of the phase space
variables. One finds immediately primary constraints: \bea
\pi^0_a(x)=0,\quad a=1,2,3. \eea The canonical Hamiltonian is now:
\bea H_c=\int
d^3x\lbrace-\frac{1}{2}\pi^{i}_{a}(x)\pi_{ia}(x)+\frac{1}{4}F_{ija}F_{ija}-A_{0a}(x)D_i\pi^{i}_{a}(x)\rbrace,
\eea with $D_\mu$ as covariant derivative: \bea
(D_\mu)_{ac}=\delta_{ac}\partial_{\mu}+\varepsilon_{abc}A_{\mu
b}(x)=\delta_{ac}\partial_{\mu}+ (I_{b})_{ac}A_{\mu b}(x). \eea The
quantities $(I_{b})_{ac}$ are elements of the adjoint representation
of the group $SU(2)$. There are also secondary constraints in the
theory , which are results of the equations of motion, namely the
conditions $\dot{\pi}^0_a(x)$=0 yield \bea
\Phi_a(x)=D_i\pi^{i}_{a}(x)=0. \eea One can check that there are no
more constraints in the theory. Using  the definition of  Poisson
brackets without taking into account any constraints gives the
following relations: \bea
\{\Phi_a(x),\Phi_b(y)\}=g\varepsilon_{abc}\Phi_c(x)\delta^3(x-y),
\eea which mean that the constraints $\Phi_a(x)$ generate local
time-independent gauge transformations. From now on we write the
Hamiltonian as follows: \bea H=H_0+\int d^3xv_a(x)\Phi_a(x), \eea in
which $H_0$ is the canonical Hamiltonian  and $v_a(x)$ are Lagrange
multipliers. In writing this we assumed   the additional conditions
(or gauge) $A_{0a}=0$,  so that the variables $A_{0a}$ and
$\pi^0_a(x)$ drop out. In general one introduces the gauge
conditions $h_a=0$ with the non-vanishing Poisson brackets of these
with $\pi^0_a(x)$ and considers the equation of motion for $h_a$.
This makes it possible to determine $A_{0a}$ as a function of the
rest of the canonical variables and we again get the Hamiltonian,
which does not depend on $A_{0a}$ and $\pi^0_a(x)$ \cite{christ}.
Nevertheless our choice of the gauge does not affect the symmetry
properties we want to study.  Thus $\Phi_a(x)$ are the only
constrains we have in the theory. In order to pass to the quantum
theory we must choose another gauge condition, define physical phase
space and corresponding Dirac brackets for independent canonical
variables. Instead we  suggest the usual Poisson bracket relations
\bea
\{A_{ia}(x),\pi^{j}_{b}(y)\}=\delta_{i}^{j}\delta_{ab}\delta^3(x-y).
\eea We then introduce Hilbert space  in which the variables
$\pi^i_a(x)$ act as differential operators $-i\delta /{\delta
A_{ia}(x)}$ and consider the constraints as the operator conditions:
\bea \Phi_a(x)\Psi=D_i\pi^{i}_{a}(x)\Psi=0, \eea in which $\Psi$ are
the vectors of Hilbert space and obey the Schr\"odinger equation.
\bea H\Psi=E\Psi. \eea In what follows we need the expressions of
conserved physical quantities as a result of  the invariance of the
equations of motion with respect to the translations and rotations
in
 3-dimensional Euclidean space or the transformations which constitute the group $M(3)$
of the motion of the space. According to the Noether's
theorem   one obtains the following expressions for the components of the momentum and  angular momentum
vectors of the system:
\bea
&P^i&=-\int d^3x \frac {\partial A_{ja}}{\partial x_i}\pi^j_a, \\
 &M_i&=\int d^3x \lbrace -(I_i)_k\,^jx_j\frac {\partial A_{na}}{\partial x_k}\pi^n_a+(I_i)_k\,^jA_{ja}\pi^k_a(x)\rbrace ,
\eea in which $(I_i)_k\,^j=\varepsilon_{ik}\,^j$ are the elements of
the adjoint representation of $SO(3)$. We construct below the
perturbation theory around the localized finite energy configuration
which takes into consideration the conservations laws of momentum
and angular momentum exactly in every order of perturbation theory.

\vspace{\baselineskip}
 {\bf  3. Collective coordinates}
\vspace{\baselineskip}

The  solutions of the equations of motion can be considered as the
functions of certain numbers of parameters, which is the result of
invariance of the equations of motion with respect to the symmetry
transformations. In accordance with this we introduce the following
transformation  of fields $A_{ia}(x)$\footnote{The matrix $C$ is a
matrix of 3-dimensional rotations. The inverse matrix to $C$ is
denoted by $\overline C$.}:
 \bea
A_{ia}(x)=C_i\,^j(\theta)[\frac{1}{g}w_{ja}(\overline {C}({\bf
x-q}))+W_{ja}({\overline C}({\bf x-q}))]. \eea We introduce with
this transformation six parameters (three rotational  $\theta_j$ and
three translational ${q_j}$, $j=1,2,3$) as new variables which
constitute together with $W_{ja}({\bf x})$ and the conjugate momenta
of the latter the  enlarged phase space. In order to retain the
original number of independent  variables we must subject them to as
many additional conditions as the number of introduced parameters.
We choose the following linear conditions: \bea \int d^3x
N_{Ik,b}\,^j({\bf x})W_{jb}({\bf x})=0,\quad I=1,2,\quad k=1,2,3,
\eea with $c$-number functions $N_{Ik,b}\,^j({\bf x})$. One finds
always the another set of $c$-number functions $M_{jb,Jk}({\bf x})$,
such that the following orthonormality conditions hold: \bea \int
d^3x N_{Ik,b}\,^j({\bf x})M_{jb,Jn}({\bf x})=\delta_{IJ}\delta_{kn}.
\eea We calculate the  variations $\delta/\delta{A_{ia}(x)}$ in the
usual way: \bea \frac{\delta}{\delta A_{ia}(x)}=\frac{\delta
q_j}{\delta A_{ia}(x)}\frac{\partial}{\partial{q_j}}+
\frac{\delta\theta_{j}}{\delta
A_{ia}(x)}\frac{\partial}{\partial{\theta_j}} +\int d^3y\frac{\delta
W_{jb}(y)}{\delta A_{ia}(x)}\frac{\delta}{\delta W_{jb}(y)}. \eea
Determining $W_{jb}(y)$ from (16)   and substituting into (17) we
then calculate the variation  of (17) with respect to  $A_{ia}({\bf
x}) $. One obtains in this way the system of algebraic equations :
\bea \overline C_j\,^iN_{Ik,a}\,^j(\overline C({\bf
x-q}))+\frac{1}{g}\tilde {N}_{Ik,a}\,^i({\bf x}) +F_{Ik,J}\,^s\tilde
{N}_{Js,a}\,^i({\bf x})=0 \eea with the functions
$\tilde{N}_{Js,a}\,^i({\bf x})$ as unknowns. The relations of the
functions $\tilde{N}_{Js,a}\,^i({\bf x})$ to the quantities we are
interested in are determined by the equalities (see the Appendix for
some notations ): \bea \frac{\delta\theta_m}{\delta A_{ia}({\bf
x})}=\overline A_m\,^k(\theta)\tilde {N}_{1k,a}\,^i({\bf x}),\qquad
\frac{\delta q_m}{\delta A_{ia}({\bf x})}={\overline C_m\,^k}{\tilde
N}_{2k,a}\,^i({\bf x}).\nn \eea The quantities $F_{Ik,Js}$  are \bea
&F_{Ik,1s}&=\int d^3xN_{Ik,b}\,^j({\bf x}) [-(\bar {I}_s)_m\,^nx_n
\frac{\partial W_{jb}({\bf x})}{\partial x_m}+(\bar {I}_s)_j\,^nW_{nb}({\bf x})], \\
&F_{Ik,2s}&=\int d^3x N_{Ik,b}\,^j({\bf x})\frac{\partial W_{jb}({\bf x})}{\partial x^s}.
\eea

Thus the quantities $\delta q_j/\delta A_{ia}(x), \,\delta\theta_{j}/\delta A_{ia}(x)$
will be determined with the use of the system of algebraic equations  and therewith
 the quantities $M_{ia,Ik}({\bf x})$ are determined, namely
\bea
M_{1s,jb}({\bf x})=(-\bar I_s)_l\,^r{x_r}\frac{\partial w_{jb}({\bf x})}{\partial x_l}+(\bar I_s)_j\,^lw_{lb}({\bf x}),\;
M_{2s,jb}({\bf x})=\frac{\partial w_{jb}({\bf x})}{\partial x^s} .\nn
\eea
In calculating  $\delta W_{jb}(y)/{\delta A_{ia}}(x)$ we have to take into account the additional
conditions (17). This can be done by defining a projection matrix
\bea
P_{ia,jb}({\bf x,y})=\delta_{ij}\delta_{ab}\delta^3({\bf x-y})-M_{ia, Ik}({\bf x})N_{I,jb}^k({\bf y}),
\eea
which has the following properties:
\bea
&&\int d^3xN_{Ik,b}\,^j({\bf x})P_{jb,ia}({\bf x,y})=\int d^3yP_{ia,jb}({\bf x,y})M_{b, Ik}^j({\bf y})=0 ,\\
\;&&\int d^3yP_{ia,jb}({\bf x,y})P^{j}_{{b},kc}({\bf
y,z})=P_{ia,kc}({\bf x,z}). \eea This operator makes it possible to
determine the variables $W_{ia}({\bf x})$(subjected to (17)) as the
linear combination of independent variables - $W_{ia}({\bf x})=\int
d^3yP_{ia,jb}({\bf x,y})V^{j}_{b}({\bf y})$ with $V_{b}^j({\bf x})$
being expressed through $A_{ia}({\bf y}),\ q_j$ and  $\theta_j$ with
the help of (16). One obtains in this way: \bea &&\frac{\delta
W_{kb}({\bf y})}{\delta A_{ia}({\bf
x})}=P_{kb,ja}({\bf y},{\overline C}({\bf x-q})) {\overline C}^{ji}+\nn \\
&&+{\tilde N}_{1s,a}\,^i({\bf x})\int d^3zP_{kb,c}\,^n({\bf y,z})
[-({\bar I}^s)_l\;^rz_r\frac{\partial W_{nc}({\bf z})}{\partial
z_l}+ ({\bar I}^s)_n\;^j W_{jc}({\bf z})] + \nn \\ &&+{\tilde
N}_{2s,a}\;^i({\bf x})\int d^3zP_{kb,c}\,^j({\bf y,z}) \frac
{\partial W_{jc}({\bf z})}{\partial z_s}. \eea We give now the final
expression for the momenta $\pi_{a}^i({\bf x})$: \bea
&&\pi_{a}^i({\bf x})={\overline
C}_j\,^i\{\Pi_a^j({\overline C}{(\bf x-q}))- \nn \\
&&-g{[\mathbb{I}+gF]^{-1}}^{sk}_{IJ} N_{Jk,a}\,^j({\overline C}({\bf
x-q}))({\bar l}_{Is}+M_{Is}(\Pi))\}. \eea The quantities
$\Pi_a^j({\bf x})$ are momenta obtained by projecting the functional
derivatives $-i\delta/{\delta W_{ia}}({\bf x})$, \bea \Pi_a^i({\bf
x})=\int d^3yP_{jb},{^{i}_{a}}({\bf
y,x})\left(-i\frac{\delta}{\delta W_{jb}({\bf x})}\right),  \nn \eea
and has the following commutation relations with the $W_{ia}({\bf
x})$: \bea [W_{ia}({\bf x}),\Pi_b^j({\bf
y})]=iP_{ia},{^{j}_{b}}({\bf x,y}). \eea Besides they satisfy the
constraint equation: \bea \int d^3x M_{ia,Ik}({\bf x})\Pi_a^i({\bf
x})=0 \eea

The dependence of momenta $\pi_{a}^i({\bf x})$ on the variables
$q_i,\;\theta_i$ are defined by the operators ${\bar l}_{Is}$. They
are generators of inverse transformations from the group $M(3)$ (see
appendix), the first three of which - $\bar l_s$ generate the
inverse rotations and others - ${\overline
C}_{ij}\partial/{i\partial q_j}$ translations in the opposite
direction (we define as forward transformations ${\bf
x}\rightarrow\overline C{\bf x+q}$). It is significant at this point
to have the conserved quantities $P^i$ and $M_i$ expressed through
the new set  of  phase space variables. On can show, that \bea
P_i=l_{1i}=-i\frac{\partial}{\partial q_i},\quad
M_i=l_{2i}=l_i-i\varepsilon_{ijk}q_j\frac{\partial}{\partial q_k},
\eea which coincide with six generators $l_{Is}=(l_{1s},l_{2s})$ of
forward transformations of the group $M(3)$. The generators  $l_i$
are generators of  $SO(3)$. These relations together with the
expression (27) are important proceeding from the statement that the
generators of  forward transformations commute with those of
backward transformations: \bea [l_{Ii},\bar l_{J,k}]=0. \eea This
means that the dependence of the state vector $\Psi$ on variables
$q_i,\, \theta_i$ can be factorized and the operators $l_{Ik}$ can
be replaced by c-number quantities. Before doing this we need to
perform a transformation of the state vector $\Psi$, namely take
\bea \Psi=\exp\left(\frac{i}{g}\int d^3xs^i_a({\bf x})W_{ia}({\bf
x})\right)\Psi', \eea which means that the momenta $\Pi^i_a ({\bf
x})$ have to be replaced by $\frac{1}{g}s^i_a({\bf x})+\Pi^i_a ({\bf
x})$. The quantities $s^i_a({\bf x})$ are subject to the same
additional conditions as $\Pi^i_a({\bf x})$: \bea
 \int d^3x M_{ia,Ik}({\bf x})s_a^i({\bf x})=0.
\eea
 These conditions can  always be
 fulfilled. If it is not so, then one defines  with the help of a projection operator
a new quantity $s'_{ia}({\bf x})$ which satisfies the needed conditions.
 After performing this transformation  one obtains for $\pi_a^i({\bf x})$:
\bea
&&\pi_{a}^i({\bf x})={\overline C}_j\,^i\{\frac{1}{g}s^j_a({\overline C}({\bf x-q}))+
\Pi_a^j({\overline C}{(\bf x-q}))-\nn \\
&&-g{[\mathbb{I}+gF]^{-1}}^{sk}_{IJ} N_{Jk,a}\,^j({\overline C}({\bf
x-q}))({\bar l}_{Is}+\frac{1}{g}M_{Is}(s)+M_{Is}(\Pi))\}. \eea This
completes the preparation of  all formulae which are needed to
construct the perturbation theory.

\vspace{\baselineskip}

{\bf 4. Quantum behaviour of localized solutions}

\vspace{\baselineskip}

 In this section we consider the Schr\"odinger
equation \bea H\Psi'=E\Psi' \eea which is supplemented by  the
additional conditions: \bea D_i\pi^{i}_{a}(x)\Psi'=0, \eea with
$\pi^{i}_{a}(x)$ as in Eq. (34). Writing out explicitly the kinetic
energy we see, that dependence of the Hamiltonian on the group
parameters is determined by generators $\bar l_{Ik}$. This fact
enables us to  separate  the dependence of the state vector on the
parameters of the symmetry group $M(3)$ in the form of a factor as
an indication of the exact fulfillment of the conservation laws,
i.e. \bea \Psi'=T(q_i,\theta_i)\Psi''((W_{ia}({\bf x}),\Pi_a^i({\bf
x})) \eea (in which $T(q_i,\theta_i)$ realizes the representation of
the group $M(3)$ \cite{vilen}) and to replace the generators $\bar
l_{Ik}$ by appropriate $c$-numbers: $\bar l_{1k}\rightarrow\bar
J_k,\;\bar l_{2i}\rightarrow(1/g^2)K_i$ with the perturbation order
of the momentum being raised, which makes it possible to see the
translational effects already in the first approximation.

The eigenvalues of the Schr\"odinger equation calculated below
perturbatively are rigorously consistent with the symmetry properties of
the physical system. Thus any approach for solving the Schr\"odinger
equation describes the symmetry properties exactly in every order of
approximation.

We can now expand the Hamiltonian and the constraint   as a series
in inverse powers of $g$. The energy and state vector also must be
expanded in appropriate series. Thus we have: \bea
&&H=g^2H_0+gH_1+H_2 +g^{-1}H_{3}+g^{-2}H_{4}+\cdots,  \\
&&E=g^2E_0+gE_1+E_2 +g^{-1}E_{3}+g^{-2}E_{4}+\cdots, \\
&&\Psi''=\Psi_0+g^{-1}\Psi_1+\cdots . \eea The leading (or zero)
approximation of the perturbation theory \bea (H_0-E_0)\Psi_0=0 \nn
\eea does not consist of field operators and is completely
determined by $c$-numbers, so the energy of zeroth order is \bea
E_0=\int d^3x \{\frac{1}{4}f_{ija}f_{ija}+\frac{1}{2}(s_{ia}({\bf
x})-N_{2j,ia}({\bf x})K^j) (s_{ia}({\bf x})-N_{2j,ia}({\bf x})K^j)\}
\eea with arbitrary nonzero $\Psi_0$. The numbers $s_{ia}({\bf x})$
remain in this order undetermined. The quantity  $f_{ija}$ is a
strength tensor, constructed by means of the classical field
solutions. The next approximation reads: \bea (H_1-E_1)\Psi_0=0.
\eea The operator $H_1$ is linear in  $W_{ia}({\bf x}),\Pi_a^i({\bf
x})$. The regularity of the function $\Psi''$ with respect to
$W_{ia}({\bf x}),\Pi_a^i({\bf x})$
 requires $H_1$ and $E_1$ to be zero identically. This leads us to the determination of the quantities
$s_{ia}({\bf x})$, namely if we choose \bea s_{ia}({\bf
x})-N_{2j,ia}({\bf x})K^j=M_{ia,2k}({\bf x})c^k, \eea then the terms
linear in  $\Pi_a^i({\bf x})$  vanish. Collecting the terms at
$W_{ia}({\bf x})$ we obtain: \bea H_1=\int d^3x
\left[c_mc_n\frac{\partial^2w_{jc}({\bf x})}{\partial x_m\partial
x_n}-\right. \left.\frac{\partial f_{ijc}}{\partial
x_i}-\varepsilon_{cba}w^i_b({\bf x})f_{ija}\right]W^{j}_{c}({\bf
x})=0. \eea We see that the terms in  front of $W_{jc}({\bf x})$
coincide with the classical Euler-Lagrange equations provided
$w_{0a}({\bf x})=0$. Thus $H_1$ is identically zero. On substituting
(43) into the expression for $E_0$ we are reassured that this is
just the energy of the classical localized configurations, provided
$w_{0a}({\bf x})=0$. We return now to the Eq. (33). On the basis of
(43) it is easy to obtain from (33) the following expression for
$K^j$: \bea K^j=c_i\int d^3x\frac{\partial w_{na}({\bf x})}{\partial
x_i} \frac{\partial w^n_a({\bf x})}{\partial x_j} \eea which are
just  components of the classical momentum.  Considering  now the
derivative of the energy $E_0$ with respect to $c_k$, we find that:
\bea c_i=g^2\frac{\partial E_0}{\partial K^i}. \eea This means, that
${\bf c}$  is a velocity of the center-of-mass of the system.

At this level  the wave function $\Psi_0 (W_{jc}({\bf
x}),\Pi_a^i({\bf x}))$ is still undetermined. The next approximation
to the ground level energy $E_0$ is determined from the terms of
order $g^0$: \bea (H_2-E_2)\Psi_0(W_{jc}({\bf x}),\Pi_a^i({\bf
x}))=0. \eea The quantity $H_2$ is a quadratic form of the operators
$W_{ia}({\bf x})$ and $\Pi_a^i({\bf x})$ and can be diagonalized
being reduced to an infinite set of the oscillators. First we
consider the constraints. From now on we can choose the system of
center-of-mass by
 fixing ${\bf c}=0$,
that is ${\bf K}=0$. This suggestion simplifies the Hamilton
operator and the constraints  (12). In this system we obtain for
$H_2$: \bea H_2=\int d^3x\left[ \frac{1}{2}\Pi_{ai}({\bf
x})\Pi_{ai}({\bf x})+W_{ib}({\bf x})\right. \left.\frac{1}{2}\hat
{O}_{ib,jc}({\bf x})W_{jc}({\bf x})\right], \eea in which \bea \hat
{O}_{ib,jc}=-\delta_{ij}(D^{cl}_kD^{clk})_{bc}+(D^{cl}_iD^{cl}_j)_{bc}+2\varepsilon_{abc}f_{ija}
\eea with \bea ({D^{cl}}_j)_{ac}=\delta_{ac}\frac{\partial}{\partial
x^j}+\varepsilon_{abc}w_{jb}({\bf x}).
 \eea

The operator $\hat {O}_{ib,jc}$ depends on the classical
configurations and, of course, the knowledge of the exact form of
such solutions is needed in order to describe the corresponding
spectrum of energy. We are interested mostly in that part of the
spectrum which arises due to the symmetry properties of the system
and therefore we perform the diagonalization of $H_2$ only formally.
Suppose the functions $U_{ia}^{(bj)}({\bf x},n)$ constitute a set of
the solutions of the following system of equations: \bea \hat
{O}_{ib,c}\,^jU_{jc}({\bf x},n)=E(n)U_{ib}({\bf x},n) \eea with the
orthonormality conditions: \bea \int d^3xU^{*}{_{ia}^{(bj)}}({\bf
x},n)U_{ia}^{(b'j')}({\bf
x},n')=\delta_{bb'}\delta_{jj'}\delta_{nn'} \eea We now consider the
expansion of the field $W_{ia}({\bf x})$ in terms of
$U_{ia}^{(bj)}({\bf x},n)$: \bea W_{ia}({\bf
x})={\sum_{n}}'\sqrt{\frac{1}{2E(n)}}\left(\alpha_{bk}(n)e_j^{(k)}U_{ia}^{(bj)}({\bf
x},n)\right.+
\left.\alpha_{bk}^{+}(n)e_j^{(k)}U^{*}{_{ia}^{(bj)}}({\bf
x},n)\right), \eea where the prime indicates  that the sum does not
contain the modes with zero energy (related to the translations and
rotations). The summation number  $n$ may in general be composed of
discrete and continuous variables. The creation and annihilation
operators $\alpha_{bj}(n),\,\alpha_{bj}^{+}(n)$ obey the commutation
relations \bea
[\alpha_{ia}(n),\,\alpha_{bj}^{+}(n')]=\delta_{ij}\delta_{ab}\delta_{nn'}.
\eea The  polarization vectors $e_j^{(k)}$ make an orthonormal set
of unit vectors in 3-dimensional Euclidean space. We  next have to
deal with the unphysical degrees of freedom related to the
constraints. Up to  order $g^0$  the constraints are: \bea
({D^{cl}}_j)_{ac}\Pi_c^j({\bf x})\Psi''=0, \eea which are linear in
field operators. The unphysical degrees of freedom related to these
constraints can be eliminated by  introducing the Coulomb gauge
conditions  written as expectation value \bea (\Psi'',\partial^i
W_{ia}({\bf x})\Psi'')=0. \eea Since (55) and (56) are linear, we
can consider the state vector obtained by acting on the vacuum with
the creation operators pertaining to the various polarizations, so
the state vector might be represented as \bea \Psi''=\Psi^t\Psi^l,
\eea where $\Psi^t$ corresponds to the transverse components of the
field $W_{ia}({\bf x})$, while $\Psi^l$ is obtained by acting on the
vacuum with longitudinal creation operators. We demand that the
positive frequency part of $\partial^i W_{ia}({\bf x})$  annihilates
the vector of Hilbert space. We are interested in examining the
consequences of gauge conditions on the state $\Psi^l$. So we may
write \bea \sum_{n}\alpha_{bk}(n)e_j^{(k)}{A_{ba}}^j\Psi^l=0, \eea
with ${A_{ab}}^j=\partial^i{U_{iba}}^j({\bf x},n)$. Since we choose
the polarization vectors orthogonal to the quantities ${A_{ab}}^j$,
(58) can be written in the equivalent form: \bea \alpha_{3a}\Psi^l=0
\eea We present the state $\Psi^l$  as the linear combination of the
eigenstates $|n>$ of the quantum number operator
$N=\sum_{n}\alpha_{3a}^{+}(n)\alpha_{3a}(n)$ corresponding to the
longitudinal degrees of freedom. The states $|n>$ obeying (59)
satisfy the condition \bea n<n|n>=0\quad {\mbox {\rm
or}}\;<n|n>=\delta_{n0} \nn \eea and therefore for the state
$\Psi^l$ as linear combination of $|n>$ the following relation
holds: $(\Psi^l,\Psi^l)=c<0|0>{\not =}0$.  Expanding the operators
in a series similar to (53) we have \bea \Pi_{ia}({\bf
x})=i{\sum_{n}}'\sqrt{\frac{E(n)}{2}}\left(\alpha_{bk}^{+}(n)e_j^{(k)}U^{*}{_{ia}^{(bj)}}({\bf
x},n)\right.- \left.\alpha_{bk}(n)e_j^{(k)}U_{ia}^{(bj)}({\bf
x},n)\right), \eea and substituting (53), (60) into (48) we obtain:
\bea
H_2=\frac{1}{2}\sum_{n}\left(\alpha_{bk}^{+}(n)\alpha_{bk}(n)+\alpha_{bk}(n)\alpha_{bk}^{+}(n)\right),
\eea which has to be understood as the normal product. Calculating
the expectation  value of $H_2$ we see that only transversal
components contribute to the energy, since the action  of $H_2$ on
the state (57) cancels the longitudinal contribution.

In order to investigate the internal structure of the system
associated with spin, we must deal with the Hamiltonian up to the
order $g^2$. Combining the terms of order $g$, that is $H_3$ , we
can show that after averaging over the ground-state wave function
$\Psi''$ its contribution to the energy is zero.  We do not give the
explicit form of $H_4$ here since this consists of  a lot of terms.
We just indicate that it is a quartic in field operators.

The expectation  value of $H_4$ gives the  contribution to the energy,
which results in a splitting  of the ground-state energy. Omitting the
 irrelevant for the structure of the energy spectrum additive terms
  we obtain for the energy above the ground level:
 \bea
 E_4=\frac{1}{2}\left(\int d^3x N_{ia,1k}({\bf x})N_{ia,1,k}({\bf x})\right)j(j+1)
\eea which  describes the excited quantum states with the spin $j$. The
 number $j$ takes either integer or half-integer values. Since the initial vector fields
 $A_{\mu a}$ belong to the adjoint representation of the group of
 rotations $j$ must be integer, hence $j=0,1,2\dots$.
 Thus, we see that there are excitation modes
 in quantum theory of the
localized finite energy solutions which are labelled by the quantum
numbers corresponding to the spin of the system.  Besides there is a
degeneracy of the energy spectrum with respect to the azimuthal
quantum number. So the energy $E_4$ defines a spin content of the
energy spectrum.

\vspace{\baselineskip}
{\bf  Acknowledgment: }A. V. S.  is indebted
to DAAD for supporting his visit to Kaiserslautern University.

\vspace{\baselineskip}

 {\bf Appendix}

 \vspace{\baselineskip}

We recall here some relevant formulae from group theory. Let the
matrices $C$ be the elements of the group $M$. We denote by
$\overline C$ the inverse  transformations. If the parameters of the
group are $\theta_i$ then the following equations hold: \bea
\frac{\partial C}{\partial\theta_i}=B^{ij}(\theta)T_jC , \nn \eea in
which $T_j$ are the elements of the corresponding algebra. The
matrix $B^{ij}(\theta)$ is invertible -
$A_{ik}(\theta)B^{kj}(\theta)=\delta_i^j$. With the matrix
$A_i\,^k(\theta)$ the generators of the group are determined: \bea
L_i=A_{ik}(\theta)\frac{\partial}{\partial\theta_k} \nn \eea with
the commutation relations \bea [L_i,L_k]=f_{ikj}L^j .\nn \eea The
inverse matrices obey the same equations, but with  quantities being
overlined: \bea \frac{\partial \overline
C}{\partial\theta_i}=\overline B^{ij}(\theta)(\overline
T_j)\overline C \nn ,\qquad \overline L_i=\bar
A_{ik}(\theta)\frac{\partial}{\partial\theta_k} \nn,\qquad
[\overline L_i,\overline L_k]=f_{ikj}\overline L^j,  \nn \eea so
that ${\overline L_i}=C_i^jL_j$, ${\overline T_i}=-T_i$ and
$[\overline L_i,L_j]=0$. The generators of the forward
transformations of the group $M(3)$ we are interested in are \bea
l_{1i}=l_i+\varepsilon_{ijk}q^j\frac{\partial}{i\partial q_k},\qquad
l_{2i}=\frac{\partial}{i\partial q^i}.\nn \eea and the inverse
transformations are \bea \bar l_{1i}=\bar l_i,\qquad \bar
l_{2i}={\overline C}_{ij}\frac{\partial}{i\partial q_j}. \nn \eea

\end{document}